\documentclass[fleqn]{article}
\usepackage[centertags]{amsmath}
\usepackage{espcrc2}
\usepackage{graphicx}
\title{Probing Confinement with Chromomagnetic Fields}

\author{Paolo Cea\address[DF,INFN]{Dipartimento di Fisica, Univ. of Bari and INFN - Sezione di Bari,
        I-70126 Bari, Italy}  and
        Leonardo Cosmai\address[INFN]{INFN - Sezione di Bari, I-70126 Bari, Italy}}

\begin{document}

\begin{abstract}
Using the lattice  Schr\"odinger functional we study vacuum
dynamics of  SU(3) gauge theory at finite temperature. The vacuum
is probed by means of an external constant Abelian  chromomagnetic
field. We find that by increasing the strength of the applied
external field the  deconfinement temperature decreases towards
zero. This implies that strong enough Abelian chromomagnetic
fields destroy confinement of color. \vspace{1pc}
\end{abstract}

\maketitle

To study vacuum structure of lattice gauge theories we
introduced~\cite{Cea:1997ff,Cea:1999gn} a gauge invariant
effective action, defined by means of the lattice Schr\"odinger
functional
\begin{equation}
\label{Zetalatt}
{\mathcal{Z}}[U^{\mathrm{ext}}_\mu] = \int {\mathcal{D}}U \; e^{-S_W} \,.
\end{equation}
$S_W$ is the standard Wilson action and the functional integration
is extended over links on a lattice $L_s^3 \times L_4$  with the
hypertorus geometry  and satisfying the constraints ($x_t$:
temporal coordinate)
\begin{equation}
\label{coldwall} U_k(x)|_{x_t=0} = U^{\mathrm{ext}}_k(x)
\,,\,\,\,\,\, (k=1,2,3) \,\,,
\end{equation}
$U^{\mathrm{ext}}_k(x)$ being the lattice version of
the external continuum gauge field
$\vec{A}^{\mathrm{ext}}(x)=
\vec{A}^{\mathrm{ext}}_a(x) \lambda_a/2$.

The lattice effective action
for the external static background field
$\vec{A}^{\mathrm{ext}}(x)$ is given by
\begin{equation}
\label{Gamma} \Gamma[\vec{A}^{\mathrm{ext}}] = -\frac{1}{L_4} \ln
\left\{
\frac{{\mathcal{Z}}[\vec{A}^{\mathrm{ext}}]}{{\mathcal{Z}}[0]}
\right\}
\end{equation}
and is invariant for  gauge transformations
of the external links $U^{\mathrm{ext}}_k$.

At finite temperature $T=1/a L_t, \, L_t \ll L_s $, we
introduced~\cite{Cea:2001an} the thermal partition function in
presence of a given static background field:
\begin{equation}
\begin{split}
\label{ZetaTnew} & \mathcal{Z}_T \left[ \vec{A}^{\text{ext}}
\right]
= \\[-0.3cm]
& \qquad \qquad
\int_{U_k(L_t,\vec{x})=U_k(0,\vec{x})=U^{\text{ext}}_k(\vec{x})}
\mathcal{D}U \, e^{-S_W}   \,.
\end{split}
\end{equation}
In this case the relevant quantity is the free energy functional
\begin{equation}
\label{freeenergy} F[\vec{A}^{\mathrm{ext}}]= - \frac{1}{L_t} \ln
\frac{\mathcal{Z}_T[\vec{A}^{\mathrm{ext}}]} {{\mathcal{Z}}_T[0]}
\,.
\end{equation}
We used the above defined lattice effective actions to investigate
the response of the vacuum to an external constant Abelian
chromomagnetic field:
\begin{equation}
\label{field}
\vec{A}^{\mathrm{ext}}_a(\vec{x}) =
\vec{A}^{\mathrm{ext}}(\vec{x}) \delta_{a,3} \,, \quad
A^{\mathrm{ext}}_k(\vec{x}) =  \delta_{k,2} x_1 H \,.
\end{equation}
On the lattice we have:
\begin{equation}
\label{t3links}
\begin{split}
& U^{\mathrm{ext}}_1(\vec{x}) =
U^{\mathrm{ext}}_3(\vec{x}) = {\mathbf{1}} \,,
\\
& U^{\mathrm{ext}}_2(\vec{x}) =
\begin{bmatrix}
\exp(i \frac {g H x_1} {2})  & 0 & 0 \\ 0 &  \exp(- i \frac {g H
x_1} {2}) & 0
\\ 0 & 0 & 1
\end{bmatrix}
\end{split}
\end{equation}
To be consistent with the hypertorus geometry we impose that the
magnetic field  is quantized:
\begin{equation}
\label{quant} \frac{a^2 g H}{2} = \frac{2 \pi}{L_1}
n_{\mathrm{ext}} \,, \qquad  n_{\mathrm{ext}}\,\,\,{\text{integer}}\,.
\end{equation}
\begin{figure}[!ht]
\begin{center}
\includegraphics[width=0.4\textwidth,clip]{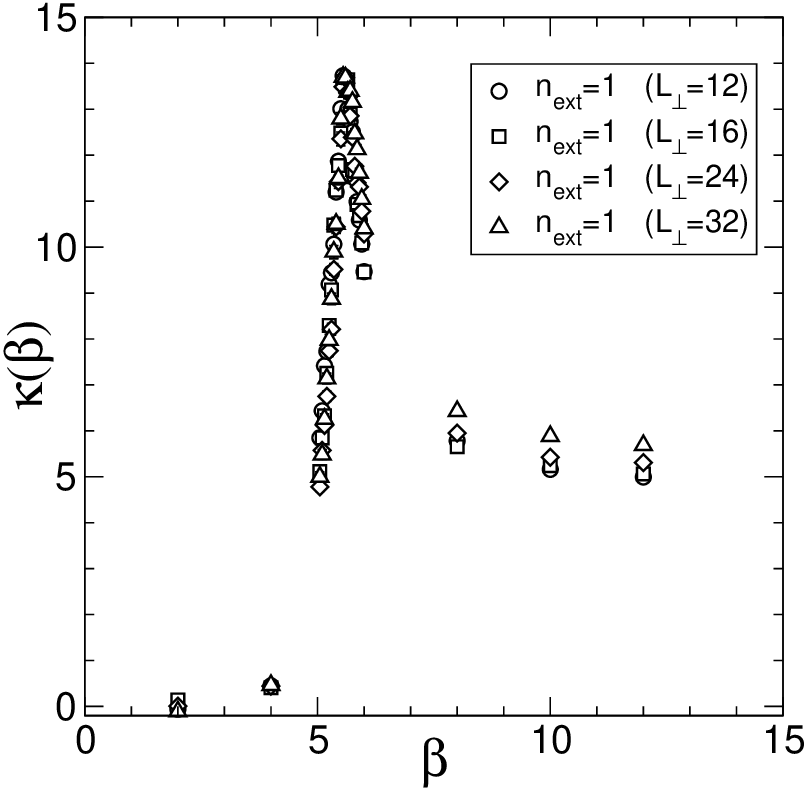}
\vspace{-1.0cm} \caption{The scaling function $\kappa(\beta)$
 versus $\beta$.}
\end{center}
\end{figure}
This field gives rise to constant field strength, then due to
gauge invariance it is easy to show that both the vacuum energy
density and the free energy density are proportional to spatial
volume $V=L_s^3$. Therefore the relevant quantity is the density
of energy $\varepsilon[\vec{A}^{\mathrm{ext}}]$ or  free energy
$f[\vec{A}^{\mathrm{ext}}]$.
At zero temperature we consider the derivative of
 $\varepsilon[\vec{A}^{\mathrm{ext}}]$ with respect to $\beta$ by
taking  $gH$ fixed and consider only the contribution due to the
dynamical links (i.e. the links that are not constrained in the
functional integration):
\begin{equation}
\label{deriv}
\begin{split}
 \varepsilon^{\prime}_{\mathrm{int}}[\vec{A}^{\mathrm{ext}}]  & =
\left \langle \frac{1}{\Omega_{\mathrm{int}}} \sum_{x \in
\tilde{\Lambda},\mu > \lambda}
\frac{1}{3} \text{Re} {\text{Tr}} U_{\mu\nu}(x) \right\rangle_0  \\
& - \left\langle \frac{1}{\Omega_{\mathrm{int}}} \sum_{x \in
\tilde{\Lambda},\mu > \lambda} \frac{1}{2} \text{Re} {\text{Tr}}
U_{\mu\nu}(x) \right\rangle_{\vec{A}^{\mathrm{ext}}} \,,
\end{split}
\end{equation}
where $\tilde{\Lambda}$ is the ensemble of the internal lattice
sites. In order to extract
$\varepsilon_{\mathrm{int}}(\beta,n_{\mathrm{ext}})$ we can
numerically integrate the data for
$\varepsilon^{\prime}_{\mathrm{int}}(\beta,n_{\mathrm{ext}})$
using the trapezoidal rule.

We find that our data satisfy the scaling law:
\begin{equation}
\label{universal}  x^{-\alpha}
\frac{\varepsilon^{\prime}_{\mathrm{int}}
(\beta,n_{\mathrm{ext}},L_{\mathrm{eff}})}
{\varepsilon^{\prime}_{\mathrm{ext}}}  = \kappa(\beta) \,.
\end{equation}
where $x = \frac{a_H}{L_{\mathrm{eff}}}$, ${a_H = \sqrt{ \frac{2
\pi}{g H}}}$ is the magnetic length, and ${L_{\mathrm{eff}} =
\Omega_{\mathrm{int}}^{1/4}}$ is the lattice effective linear size
(see Fig.~1). In Eq.~(\ref{universal})
$\varepsilon^{\prime}_{\mathrm{ext}}$ is the  derivative of the
classical energy due to  the external applied field
$\varepsilon^{\prime}_{\mathrm{ext}} = \frac{2}{3} \, [1 - \cos(
\frac{g H}{2} )] = \frac{2}{3} \, [1 - \cos( \frac{2 \pi}{L_1}
n_{\mathrm{ext}})]$ .
%
%
%
Remarkably, the value of the exponent $\alpha = 1.5$ in
Eq.~(\ref{universal}) agrees with the one we found for the SU(2)
gauge theory. From Eq.~(\ref{universal}) we can determine the
infinite volume limit of the vacuum energy density
$\varepsilon_{\mathrm{int}}$:
\begin{equation}
\label{infinite-vol}
 \lim_{L_{\mathrm{eff}} \to \infty}
\varepsilon_{\mathrm{int}}(\beta,n_{\mathrm{ext}},L_{\mathrm{eff}})=
0 \; .
\end{equation}
As a consequence, in the continuum limit the SU(3) vacuum screens
the external chromomagnetic Abelian field. \\
At finite temperature the relevant quantity is the
$\beta$-derivative of $f[\vec{A}^{\mathrm{ext}}]$:
\begin{equation}
\label{derivT}
\begin{split}
f^{\prime}[\vec{A}^{\mathrm{ext}}]   & = \left \langle
\frac{1}{\Omega_{\mathrm{int}}} \sum_{x \in \tilde{\Lambda},\mu <
\nu}
\frac{1}{3} \text{Re} {\text{Tr}} U_{\mu\nu}(x) \right\rangle_0  \\
& - \left\langle \frac{1}{\Omega_{\mathrm{int}}} \sum_{x \in
\tilde{\Lambda},\mu < \nu} \frac{1}{3} \text{Re} {\text{Tr}}
U_{\mu\nu}(x) \right\rangle_{\vec{A}^{\mathrm{ext}}} \,.
\end{split}
\end{equation}
It turns out  that, if we start with the SU(3) gauge system at
zero temperature in a constant Abelian chromomagnetic background
field of fixed strength  and increase the temperature, then  the
perturbative tail of the $\beta$-derivative of the free energy
density increases with $1/L_t$ and tends towards the ``classical''
value. So that, we may  conclude that by increasing the
temperature there is no screening effect in the free energy
density confirming that the zero-temperature screening of the
external field is related to the confinement. Moreover,
$f^{\prime}_{\mathrm{int}}(\beta,n_{\mathrm{ext}})$ can be used to
get an estimate of the deconfinement temperature $T_c$. To this
end, we evaluate $f^{\prime}[\vec{A}^{\mathrm{ext}}]$ as a
function of $\beta$ for different lattice temporal sizes $L_t$. To
determine the pseudocritical couplings we parameterize
$f^{\prime}(\beta,L_t)$
 near the peak as
\begin{equation}
\label{peak-form}
\frac{f^{\prime}(\beta,L_t)}{\varepsilon^{\prime}_{\mathrm{ext}}}
= \frac{a_1(L_t)}{a_2(L_t) [\beta - \beta^*(L_t)]^2 +1} \,.
\end{equation}
Once determined $\beta^*(L_t)$ we estimate the deconfinement
temperature as
\begin{equation}
\label{Tc} \frac{T_c}{\Lambda_{\mathrm{latt}}} = \frac{1}{L_t}
\frac{1}{f_{SU(3)}(\beta^*(L_t))} \,,
\end{equation}
with
\begin{equation}
\label{af} f_{SU(N)}(\beta) = \left( \frac{\beta}{2  N b_0}
\right)^{b_1/2b_0^2} \, \exp \left( -\beta \frac{1}{4 N b_0}
\right) \,,
\end{equation}
where $N$ is the color number, $b_0=(11 N)/(48 \pi^2)$, $b_1=(34N^2)/(3(16\pi^2)^2)$.

Following~\cite{Fingberg:1993ju} we perform a linear extrapolation
to the continuum of our data for $T_c/\Lambda_{\mathrm{latt}}$
(see Fig.~2). We see that the continuum limit deconfinement
critical temperature does depend on the applied Abelian
chromomagnetic field.
\begin{figure}[!ht]
\begin{center}
\includegraphics[width=0.4\textwidth,clip]{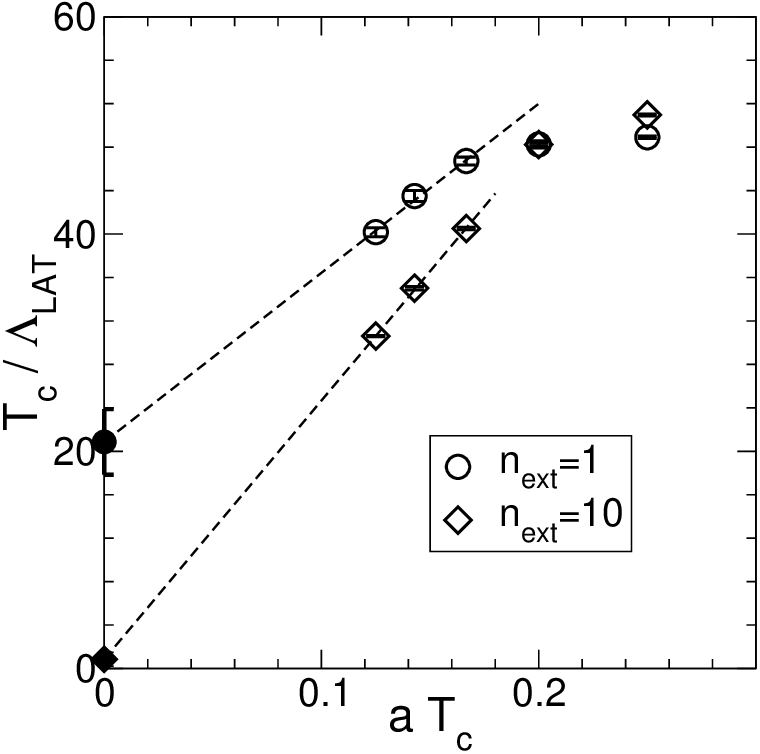}
\vspace{-1.0cm} \caption{$T_c/\Lambda_{\mathrm{latt}}$ versus
$aT_c$ for two different values of the external field strength.}
\end{center}
\end{figure}
Therefore we decided to vary the strength of the applied external
Abelian chromomagnetic background field to study quantitatively
the dependence of $T_c$ on $gH$.
%
%
%
In Fig.~3 we display our determination of $T_c$ as a function of
the external chromomagnetic field $gH$. We see that the critical
temperature decreases by increasing the external Abelian
chromomagnetic field. If the magnetic length $a_H \sim
1/\sqrt{gH}$  is the only relevant scale of the problem, then for
dimensional reasons one expects that $ T_c^2  \, \sim   \, gH$ .
%
%
%
\begin{figure}[!ht]
\begin{center}
\includegraphics[width=0.4\textwidth,clip]{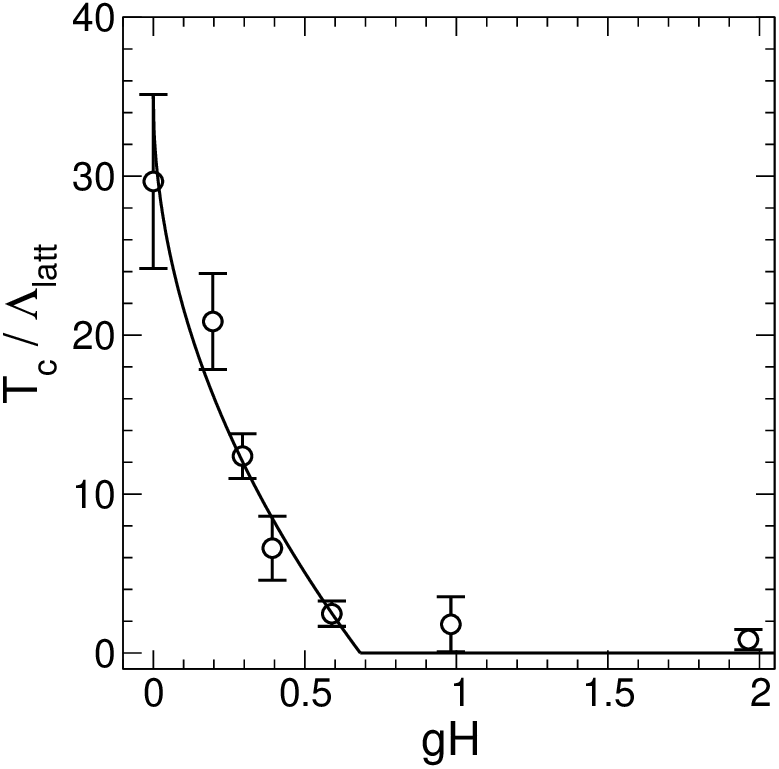}
\vspace{-1.0cm} \caption{The continuum critical temperature
$T_c/\Lambda_{\mathrm{latt}}$ versus the external field strength
$gH$. Solid line is the fit of Eq.~(\ref{Tcfit}) to our data.}
\end{center}
\end{figure}
Indeed we try to fit our data with
\begin{equation}
\label{Tcfit} \frac{T_c(gH)}{\Lambda_{\text{latt}}} =
\frac{T_c(0)}{\Lambda_{\text{latt}}} + \alpha \sqrt{gH} \,.
\end{equation}
We find a satisfying fit with $\alpha=-42.4 \pm 7.4$ and
$T_c(0)/\Lambda_{\text{latt}}$ that agrees within errors with
Ref.~\cite{Fingberg:1993ju}. We see that there exists a critical
field $gH_c \simeq 0.68$
%
such that $T_c=0$ for $gH > gH_c$. \\
The existence of a critical chromomagnetic field is naturally
explained if the vacuum behaves as an ordinary relativistic color
superconductor, a condensate of a color charged scalar field whose
mass is proportional to the inverse of the magnetic length. A
natural candidate for such  tachyonic color charged scalar field
is the famous Nielsen-Olesen unstable mode. However, it must be
observed that the  chromomagnetic  condensate cannot be uniform
due to gauge invariance of the vacuum, which disorders the system
in such a way that there are not long range color correlations. \\
Another important aspect of this work might be related to
 astrophysics  applications.  Our results implies
the exciting possibility of a  new class of compact quark stars.


\end{document}